\documentclass[aps,pra,twocolumn,showpacs,floatfix]{revtex4-1}
\usepackage{graphicx,amsfonts,amssymb,amsmath,hyperref}
\usepackage{textcomp}
\usepackage{esint}

\newif\ifhyper
\hypertrue
\ifhyper       
\hypersetup{         
  citecolor = {green},
  colorlinks = {true}, 
  urlcolor = {blue} 
}

\begin{document}

\def\rhoeq{\hat\rho_{\rm eq}}

\newcommand{\marge}[1]{\marginpar{\scriptsize #1}}
\newcommand{\remarque}[1]{\marginpar{\scriptsize Remarque}{\it [#1]}}
\newcommand{\new}[1]{{\bf #1}}
\newlength{\textlarg}
\newcommand{\barre}[1]{%
   \settowidth{\textlarg}{#1}
   #1\hspace{-\textlarg}\rule[0.5ex]{\textlarg}{0.5pt}}
\newcommand{\barred}[1]{%
   \settowidth{\textlarg}{#1}
   \red{#1}\hspace{-\textlarg}\rule[0.5ex]{\textlarg}{0.5pt}}
\newcommand{\barblue}[1]{%
   \settowidth{\textlarg}{#1}
   \blue{#1}\hspace{-\textlarg}\rule[0.5ex]{\textlarg}{0.5pt}}

\def\beq{\begin{equation}}
\def\eeq{\end{equation}}
\def\bleq{\begin{eqnarray}}
\def\eleq{\end{eqnarray}} 
\def\bfig{\begin{figure}}
\def\efig{\end{figure}}
\def\bline{\begin{multline}}
\def\eline{\end{multline}}
\def\bremark{\begin{quotation} \noindent \small }
\def\eremark{\end{quotation}}
\def\llbrace{\left\lbrace}
\def\rrbrace{\right\rbrace}
\def\lbraket{\left[}
\def\rbraket{\right]}
\def\llangle{\left\langle}
\def\rrangle{\right\rangle} 

\newcommand{\Tr}{{\rm Tr}} 
\newcommand{\tr}{{\rm tr}} 
\newcommand{\sgn}{{\rm sgn}} 
\newcommand{\mean}[1]{\langle #1 \rangle}
\newcommand{\commu}[2]{[#1,#2]} 
\newcommand{\bra}[1]{\langle#1|}
\newcommand{\ket}[1]{|#1\rangle}
\newcommand{\braket}[2]{\langle #1|#2\rangle}
\newcommand{\dbraket}[3]{\langle #1|#2|#3\rangle}
\newcommand{\tens}[1]{\overleftrightarrow{#1}}  
\newcommand{\vac}{|{\rm vac}\rangle} 
\def\bravac{\langle{\rm vac}|}
\newcommand{\const}{{\rm const}} 
\newcommand{\atanh}{\,{\rm atanh}}

\newcommand{\ie}{i.e. }
\newcommand{\iet}{i.e.}
\newcommand{\eg}{e.g. }
\newcommand{\cc}{{\rm c.c.}} 
\newcommand{\hc}{{\rm h.c.}} 
\def\etal{{\it et al. }}

\newcommand{\jhatbf}{\hat {\textbf \j}} 
\newcommand{\Jhatbf}{\hat {\textbf \J}} 
\newcommand{\jhat}{\hat {\jmath}} 
\newcommand{\Jhat}{\hat {J}} 
\newcommand{\jbf}{\textbf j}
\newcommand{\Jbf}{\textbf J}

\def\chibf{\boldsymbol{\chi}}
\def\down{\downarrow}
\def\eps{\epsilon}
\def\gam{\gamma} 
\def\phibf{\boldsymbol{\phi}}
\def\varphibf{\boldsymbol{\varphi}}
\def\varphibfs{\boldsymbol{\varphi}_<}
\def\varphibfl{\boldsymbol{\varphi}_>}
\def\varphis{\varphi_{<}}
\def\varphil{\varphi_{>}}
\def\psibf{\boldsymbol{\psi}}
\def\Ome{\Omega}
\def\omeD{{\omega_D}} 
\def\bfOme{\boldsymbol{\Omega}} 
\def\Omebf{\boldsymbol{\Omega}} 
\def\lamb{\lambda}
\def\Lamb{\Lambda}
\def\sig{\sigma}
\def\Sig{\Sigma}
\def\sigp{{\sigma'}} 
\def\bfsig{\boldsymbol{\sigma}} 
\def\sigbf{\boldsymbol{\sigma}} 
\def\The{\Theta} 
\def\up{\uparrow}

\def\epsk{\epsilon_{\bf k}} 
\def\xik{\xi_{\bf k}} 
\def\txik{\tilde\xi_{\bf k}} 
\def\xip{\xi_{\bf p}} 
\def\xiq{\xi_{\bf q}} 
\def\xikq{\xi_{{\bf k}+{\bf q}}} 
\def\Ek{E_{\bf k}} 
\def\Ep{E_{\bf p}}
\def\Eq{E_{\bf q}}
\def\Heff{\hat H_{\rm eff}}
\def\Hem{\hat H_{\rm em}}
\def\Hint{\hat H_{\rm int}}
\def\Hloc{\hat H_{\rm loc}}
\def\HMF{\hat H_{\rm MF}}
\def\Sem{S_{\rm em}}
\def\SMF{S_{\rm MF}} 
\def\SHF{S_{\rm HF}} 
\def\SRPA{S_{\rm RPA}} 
\def\Sint{S_{\rm int}} 
\def\Sloc{S_{\rm loc}}
\def\TN{T_{\rm N}} 
\def\TNHF{T^{\rm HF}_{\rm N}} 
\def\Zloc{Z_{\rm loc}} 
\def\ZMF{Z_{\rm MF}} 
\def\ZHF{Z_{\rm HF}} 
\def\ZRPA{Z_{\rm RPA}} 
\def\RPA{{\rm RPA}}
\def\loc{{\rm loc}} 
\def\pp{{\rm pp}}
\def\ph{{\rm ph}} 
\def\ch{{\rm ch}}
\def\sp{{\rm sp}} 
\def\qtf{q_{\rm TF}}
\def\epstf{\eps^{}_{\rm TF}} 
\def\epsrpa{\eps^{}_{\rm RPA}} 
\def\chinnzpp{\chi_{nn}^{0}{}\!\!\!''}

\def\half{\frac{1}{2}}
\def\dhalf{\dfrac{1}{2}}
\def\third{\frac{1}{3}} 
\def\quarter{\frac{1}{4}}

\def\qr{{\bf q}\cdot{\bf r}}
\def\wt{\omega t} 

\def\a{{\bf a}}
\def\b{{\bf b}}
\def\e{{\bf e}}
\def\f{{\bf f}}
\def\g{{\bf g}}
\def\h{{\bf h}}
\def\k{{\bf k}}
\def\l{{\bf l}}
\def\m{{\bf m}}
\def\n{{\bf n}} 
\def\p{{\bf p}} 
\def\q{{\bf q}}
\def\r{{\bf r}}
\def\t{{\bf t}}
\def\u{{\bf u}}
\def\v{{\bf v}}
\def\x{{\bf x}}
\def\y{{\bf y}} 
\def\z{{\bf z}} 
\def\A{{\bf A}}
\def\B{{\bf B}}
\def\D{{\bf D}} 
\def\E{{\bf E}} 
\def\F{{\bf F}} 
\def\H{{\bf H}}  
\def\J{{\bf J}}
\def\K{{\bf K}} 

\def\G{{\bf G}}
\def\L{{\bf L}}
\def\M{{\bf M}}  
\def\O{{\bf O}} 
\def\P{{\bf P}} 
\def\Q{{\bf Q}} 
\def\R{{\bf R}}
\def\S{{\bf S}}
\def\epsbf{\boldsymbol{\epsilon}}
\def\mubf{\boldsymbol{\mu}}
\def\nablabf{\boldsymbol{\nabla}}
\def\rhobf{\boldsymbol{\rho}}
\def\sigmabf{\boldsymbol{\sigma}} 
\def\Pibf{\boldsymbol{\Pi}}
\def\pibf{\boldsymbol{\pi}}

\def\para{\parallel}
\def\kpara{{k_\parallel}}
\def\kperp{{k_\perp}} 
\def\kperpp{{k_\perp'}} 
\def\qperp{{q_\perp}} 
\def\tperp{{t_\perp}} 

\def\w{\omega}
\def\wn{\omega_n}
\def\wm{\omega_m}
\def\wnu{\omega_\nu}
\def\wp{\omega_p} 
\def\dmu{{\partial_\mu}}
\def\dl{{\partial_l}}  
\def\dt{\partial_t} 
\def\tdt{\tilde\partial_t}
\def\dk{\partial_k}
\def\tdk{\tilde\partial_k}
\def\dx{\partial_x}
\def\dy{\partial_y} 
\def\dtau{{\partial_\tau}}  
\def\det{{\rm det}} 
\def\Pf{{\rm Pf}}

\def\dsum{\displaystyle \sum}
\def\dint{\displaystyle \int} 
\def\intt{\int_{-\infty}^\infty dt} 
\def\inttp{\int_{-\infty}^\infty dt'} 
\def\intk{\int_{\bf k}} 
\def\intkd{\int \frac{d^dk}{(2\pi)^d}}
\def\intq{\int_{\bf q}} 
\def\intr{\int d^dr}  
\def\dintr{\displaystyle \int d^dr} 
\def\intrp{\int d^dr'}
\def\dinttau{\displaystyle \int_0^\beta d\tau}
\def\dinttaup{\displaystyle \int_0^\beta d\tau'}
\def\inttau{\int_0^\beta d\tau}
\def\inttaup{\int_0^\beta d\tau'}
\def\intx{\int d^{d+1}x} 
\def\inttaur{\int_0^\beta d\tau \int d^dr}
\def\intinf{\int_{-\infty}^\infty}
\def\dinttaur{\displaystyle \int_0^\beta d\tau \int d^dr}
\def\dintinf{\displaystyle \int_{-\infty}^\infty}
\def\intw{\int_{-\infty}^\infty \frac{d\w}{2\pi}}
\def\sumr{\sum_{\bf r}} 

\def\calA{{\cal A}}
\def\calB{{\cal B}} 
\def\calC{{\cal C}} 
\def\dt{\partial_t}
\def\calD{{\cal D}}
\def\calF{{\cal F}} 
\def\calG{{\cal G}}
\def\calH{{\cal H}}
\def\calI{{\cal I}}
\def\calJ{{\cal J}}
\def\calK{{\cal K}}
\def\calL{{\cal L}} 
\def\calN{{\cal N}}
\def\calO{{\cal O}}
\def\calP{{\cal P}}  
\def\calR{{\cal R}} 
\def\calS{{\cal S}}
\def\calT{{\cal T}}
\def\calU{{\cal U}}
\def\calX{{\cal X}} 
\def\calY{{\cal Y}} 
\def\calZ{{\cal Z}} 

\def\calFbf{{\bf F}}

\def\tT{{\tilde T}}
\def\talpha{{\tilde\alpha}}
\def\tdelta{{\tilde\delta}}
\def\teta{{\tilde\eta}} 
\def\tlamb{{\tilde\lambda}}
\def\tmu{{\tilde\mu}}
\def\tphibf{{\tilde\phibf}}
\def\trho{{\tilde\rho}}
\def\tvarphibf{{\tilde\varphibf}} 
\def\tw{{\tilde\omega}}
\def\twn{{\tilde\omega_n}}

\def\asinh{{\rm asinh}} 
\graphicspath{{./figures/}}

\def\TcMF{T_c^{\rm MF}} 

\title{Critical region of the superfluid transition in the BCS-BEC crossover}

\author{T. Debelhoir}
\author{N. Dupuis}
\affiliation{Laboratoire de Physique Th\'eorique de la Mati\`ere Condens\'ee, 
CNRS UMR 7600, Universit\'e Pierre et Marie Curie, 4 Place Jussieu, 
75252 Paris Cedex 05, France}

\date{January 27, 2016} 

\begin{abstract}
We determine the size of the critical region of the superfluid transition in the BCS-BEC crossover of a three-dimensional fermion gas, using a renormalization-group approach to a bosonic theory of pairing fluctuations. For the unitary Fermi gas, we find a sizable critical region $[T_G^-,T_G^+]$, of order $T_c$,  around the transition temperature $T_c$ with a pronounced asymmetry: $|T_G^+-T_c|/|T_G^--T_c|\sim8$. The critical region is strongly suppressed on the BCS side of the crossover but remains important on the BEC side.
\end{abstract}
\pacs{74.20.-z,03.75.Ss,03.75.Hh,67.85.-d}
\maketitle

\section{Introduction}
\label{sec_intro} 

Superfluidity has been well understood for a long time in two different limiting cases. The first one is the Bardeen-Cooper-Schrieffer (BCS) superfluidity of fermions~\cite{Bardeen57}, the second one is the Bose-Einstein condensation (BEC) of bosons~\cite{Bogoliubov47}. In the BCS limit, a weak attractive interaction in a highly degenerate system of fermions induces a pairing instability. Pairs form and condensate at the same temperature $T_c$ that is orders of magnitude smaller than the Fermi energy $E_F$. In the BEC limit, superfluidity arises from condensation of bosonic atoms into a single quantum state. The internal structure of the bosonic atoms (made up of an even number of fermionic constituents) is irrelevant at temperatures of the order of the BEC temperature. 

While most of the superfluids discovered in the 20th century are firmly in one or the other limit, recently discovered materials such as high-$T_c$ superconductors require one to understand superfluidity beyond the standard paradigms. From an experimental point of view, this has become possible in cold atomic gases where Feshbach resonances allow us to tune the attractive interaction between atoms and span the entire BCS-BEC crossover~\cite{Giorgini08}. Deep on the BCS side of the resonance, where the $s$-wave scattering length $a$ between fermionic atoms in two different hyperfine states is negative and satisfies $1/k_F|a|\gg 1$ (with $n=k_F^3/3\pi^2$ the density of particles), superfluidity is well described by BCS 
theory. Deep on the BEC side, where $a>0$ and $1/k_Fa\gg 1$, superfluidity arises from BEC of tightly bound fermion pairs of size $a$. The fermion gas close to unitarity ($|a|\to\infty$) is a paradigmatic example of strongly correlated systems that exhibits several remarkable properties (for reviews, see Refs.~\cite{Chen05,Bloch08,Giorgini08,Zwerger12,Randeria14}). In particular, it shows the highest ratio $T_c/E_F\simeq 0.15$ ever observed in a fermionic superfluid (with $E_F=k_F^2/2m=(3\pi^2 n)^{2/3}/2m$)~\cite{note8}. (We set $\hbar=k_B=1$ throughout the paper.)

An unsolved issue relates to the behavior of the unitary Fermi gas in the normal (nonsuperfluid) phase. In the BCS limit, above the transition temperature the Fermi gas is a Fermi liquid. On the BEC side, the normal phase is a Bose gas of fermion-fermion dimers which dissociate only at the pairing temperature $T^*$; when $T_c\leq T\ll T^*$ the fermionic spectral function exhibits a pseudogap at low energies and Fermi-liquid behavior is suppressed. The importance of pairing fluctuations and ``preformed'' (metastable) pairs at unitarity as well as the fate of pseudogap~\cite{Stajic04,Perali04} and Fermi-liquid behavior is still unsettled~\cite{Perali11,Nascimbene11}. 
On the experimental side, the situation is unclear. Radiofrequency 
spectroscopy has been interpreted as evidence for a pseudogap state~\cite{Gaebler10} and the small shear viscosity  suggests the absence of quasiparticles above $T_c$~\cite{Cao11}. On the other hand, Fermi-liquid behavior was observed in thermodynamics~\cite{Nascimbene10,Navon10} and spin transport properties~\cite{Sommer11} in contradiction with the existence of a pseudogap. 

In this paper, we study pairing fluctuations in the BCS-BEC crossover and determine the size of the critical region about the superfluid transition where thermodynamic quantities show critical behavior with critical exponents corresponding to the Wilson-Fisher fixed point of the three-dimensional classical O(2) model. In weak coupling superconductors, due to their large $T=0$ coherence length, the critical region is too small to be observed experimentally; its size scales with $(T_c/E_F)^4$. For high-$T_c$ superconductors, it can reach $10^{-2}$. It has previously been pointed out that the size of the critical region in the unitary Fermi gas should be of order unity according to the Ginzburg criterion~\cite{Taylor09}. However, a precise determination of the critical region requires a good description of critical pairing fluctuations and is therefore beyond the reach of standard theoretical approaches such as Nozi\`eres--Schmitt-Rink (NSR) theory or $T$-matrix approximation~\cite{Nozieres85,Sademelo93,Chen05}.
The superfluid transition temperature of the unitary Fermi gas has been determined accurately by a diagrammatic determinant Monte Carlo method~\cite{Burovski08} but the size of the critical region has not been estimated. 
We have used the nonperturbative renormalization group (NPRG) to obtain an estimate of the critical region of the unitary Fermi gas~\cite{[{The NPRG approach has been used previously to determine the critical region in quantum Ising models; see }] Jakubczyk08a,*Strack09}. Our approach differs from previous NPRG studies (for reviews, see Refs.~\cite{Scherer10,Boettcher12}) insofar as we integrate out the fermions using a complex Hubbard-Stratonovich field and deal with a bosonic theory of pairing fluctuations (for a perturbative RG approach along the same lines, see~\cite{Gubbels11}). 

We compute the Ginzburg temperatures $T_G^+$ and $T_G^-$ of the unitary Fermi gas and find a sizable critical region $[T_G^-,T_G^+]$ around the transition temperature $T_c$, of the order of $T_c$, with a pronounced asymmetry: $|T_G^+-T_c|/|T_G^--T_c|\sim8$.
The critical region in the BCS-BEC crossover is estimated from a Ginzburg criterion with the additional constraint that the NPRG results should be reproduced in the unitary limit. We find that the critical region is strongly suppressed on the BCS side of the crossover but remains important on the BEC side.

\section{NPRG approach} 
\label{sec_critical} 

We consider a system of neutral fermions with two possible hyperfine states and dispersion $\eps_\q=\q^2/2m$, and Hamiltonian
\begin{equation}
\hat H = \int d^3r \biggl\lbrace \sum_{\sig=\up,\down} \hat\psi^\dagger_\sig \left( - \frac{\nablabf^2}{2m} - \mu \right) \hat\psi_\sig - g \hat\psi^\dagger_\up \hat\psi^\dagger_\down \hat\psi_\down \hat\psi_\up \biggr\rbrace 
\end{equation}
where the chemical potential $\mu$ controls the density of particles. We take $g>0$ so that the interaction between fermions is attractive. The $s$-wave scattering length is then defined by 
\begin{equation}
\frac{m}{4\pi a} = - \frac{1}{g} + \int_{|\q|\leq\Lambda_F} \frac{d^3q}{(2\pi)^3} \frac{1}{2\eps_\q} , 
\label{adef} 
\end{equation}
where the ultraviolet momentum $\Lambda_F$ on the fermion dispersion is introduced to regularize the zero-range interaction potential $g\delta(\r-\r')$. We set $\hbar=k_B=1$ throughout the paper. 

The partition function $Z=\Tr\,e^{-\beta\hat H}$ can be written as a functional integral over anticommuting Grassmann fields with (Euclidean) action 
\begin{equation}
S=\inttau\biggl\lbrace\int d^3r  \sum_\sig\psi^*_\sig\dtau \psi_\sig + H[\psi^*,\psi] \biggr\rbrace. 
\end{equation}
Decoupling the interaction term in the particle-particle channel using a complex Hubbard-Stratonovich field $\varphi$ and integrating out the fermion field $\psi$, we obtain the action 
\begin{equation}
S[\varphi^*,\varphi] = \frac{1}{g} \inttau \int d^3r |\varphi(x)|^2 - \Tr \ln (\calG_0^{-1} + \hat\varphi),
\label{action3} 
\end{equation}
where $\beta=1/T$ and $x=(\r,\tau)$.
\begin{equation}
\calG_0^{-1}(\q,i\wn) = \left( 
\begin{array}{cc} 
 i\wn - \eps_\q+\mu & 0 \\ 
 0 & i\wn + \eps_\q-\mu  
\end{array}
\right)
\end{equation}
and 
\begin{equation}
\hat\varphi(x,x') = \delta(x-x') \left( 
\begin{array}{cc} 
 0 & \varphi(x) \\ 
 \varphi(x)^* & 0 
\end{array}
\right)
\end{equation}
are $2\times 2$ matrices in the Nambu space defined by the two-component spinor $\Psi=(\psi_\up,\psi_\down)^T$. Here $\calG_0(\q,i\wn)$ denotes the Fourier transform of $\calG_0(x,x')$ with $\wn=(2n+1)\pi T$ ($n$ integer) a fermionic Matsubara frequency.  

The action $S[\varphi^*,\varphi]$ contains all information about pairing fluctuations and criticality. A saddle-point computation of the partition function $Z=\int\calD[\varphi^*,\varphi]e^{-S[\varphi^*,\varphi]}$ reproduces the BCS mean-field theory. Including Gaussian fluctuations about the saddle-point solution yields the NSR approach~\cite{Nozieres85}. To take into account fluctuations beyond the perturbative approach, we use the NPRG~\cite{Berges02,Delamotte12,Kopietz_book}. We add to the action the infrared regulator term 
\begin{equation}
S^{\rm reg}_k[\varphi^*,\varphi] = \sum_{\p,\wm} \varphi^*(\p,i\wm) R_k(\p,i\wm) \varphi(\p,i\wm) 
\label{action1} 
\end{equation}
($\wm=2m\pi T$ is a bosonic Matsubara frequency) 
indexed by a momentum scale $k$ such that fluctuations are smoothly taken into account as $k$ is lowered from a microscopic scale $\Lambda$ down to zero. 

The central quantity in the NPRG approach is the scale-dependent effective action (or Gibbs free energy)
\begin{align}
\Gamma_k[\Delta^*,\Delta] ={}& - \ln Z_k[J^*,J] + \inttau \int d^3r (J^*\Delta+\Delta^* J) 
\nonumber \\ & - S^{\rm reg}_k[\Delta^*,\Delta]
\end{align} 
defined as a (slightly modified) Legendre transform of the (scale-dependent) free energy $-\ln Z_k[J^*,J]$ that includes the subtraction of $S^{\rm reg}_k[\Delta^*,\Delta]$. Here $J$ is a complex external source that couples linearly to the pairing field $\varphi$ and $\Delta(x)=\mean{\varphi(x)}$ is the order parameter. The cutoff function $R_\Lambda(p)$ at the microscopic scale $\Lambda$ is chosen very large so that it suppresses all fluctuations and the mean-field theory becomes exact: $\Gamma_\Lambda[\Delta^*,\Delta]=S[\Delta^*,\Delta]$. For a generic value of $k$, $R_k(p)$ suppresses fluctuations with $|p|,|\wm|/c\lesssim k$ but leaves unaffected those with $|p|,|\wm|/c\gtrsim k$ where $c$ is the velocity of low-energy excitations. The effective action of the original model~(\ref{action3}) is given by $\Gamma_{k=0}$ provided that $R_{k=0}$ vanishes. The variation of the effective action with $k$ is governed by Wetterich's equation~\cite{Wetterich93},
\begin{equation}
\dk \Gamma_k[\Delta^*,\Delta] = \half \Tr \llbrace \frac{\partial R_k}{\partial k} \bigl(\Gamma^{(2)}_k[\Delta^*,\Delta]+R_k\bigr)^{-1} \rrbrace , 
\label{floweq}
\end{equation}
where $\Gamma_k^{(2)}$ denotes the second-order functional derivative of $\Gamma_k$. Thus the NPRG approach aims at finding $\Gamma_{k=0}$ using Eq.~(\ref{floweq}) starting from the initial condition $\Gamma_\Lambda$. 

At momentum scale $\Lambda$, all physical quantities of interest can be obtained from $\Gamma_\Lambda[\Delta^*,\Delta]=S[\Delta^*,\Delta]$. In particular, the state of the system (normal or superfluid) is determined by the constant (i.e., uniform and time-independent) minimum $\Delta_\Lambda$ of $\Gamma_\Lambda$. 
At unitarity, $\Delta_{\Lambda}$ becomes nonzero below the mean-field transition temperature $\TcMF\simeq 0.5E_F$. In the following, we want to study the vicinity of the (true) transition temperature $T_c\simeq 0.15 E_F$ which is much smaller than $\TcMF$~\cite{Burovski08}. This implies that $\Delta_{\Lambda}$  is a large energy scale of the order of $\Delta_\Lambda(T=0)\simeq 0.69E_F$~\cite{note5}.
Restricting ourselves to fluctuations satisfying $|\p|,|\w|/c_\Lambda\lesssim \Delta_\Lambda$ (with $c_\Lambda$ the Bogoliubov-Anderson mode velocity at mean-field level), we can expand the effective action $\Gamma_\Lambda$ to second order in derivatives. Assuming a similar derivative expansion for all values of $k$ smaller than $\Lambda$, this leads us to the ansatz 
\begin{align}
\Gamma_k[\Delta^*,\Delta] ={}& \inttau \int d^3r \bigl\lbrace \Delta^*(Z_{C,k} \dtau - V_{A,k} \partial_\tau^2 \nonumber \\ &- Z_{A,k} \nablabf^2) \Delta + U_k(\rho) \bigr\rbrace  
\label{action2} 
\end{align}
for the scale-dependent effective action, where the effective potential $U_k$ is a function of the U(1) invariant $\rho=|\Delta|^2$. We further expand $U_k$ to quartic order in the field about the equilibrium state $\rho=\rho_{0,k}$ defined by the minimum of $U_k$, 
\begin{equation}
U_k(\rho) = U_k(\rho_{0,k}) + \delta_k (\rho-\rho_{0,k}) + \frac{\lambda_k}{2} (\rho-\rho_{0,k})^2 ,
\end{equation}
where $\delta_k=U_k'(\rho_{0,k})$ vanishes if $\rho_{0,k}>0$. 

Since the derivative expansion~(\ref{action2}) is appropriate only for low-energy fluctuations satisfying $|\p|,|\wm|/c_\Lambda\lesssim \Delta_\Lambda/c_\Lambda$, the initial momentum cutoff $\Lambda$ must be of the order of $\Delta_\Lambda/c_\Lambda$. We thus assume that high-energy fluctuations do not significantly  alter the transition temperature and the size of the critical region. The precise value of the cutoff $\Lambda$ is fixed so as to reproduce the known transition temperature $T_c\simeq 0.15 E_F$ of the unitary Fermi gas~\cite{Burovski08}, and internal consistency of our approach requires $\Lambda\sim\Delta_\Lambda/c_\Lambda$.

The initial values $\delta_\Lambda$ and $\lambda_\Lambda$ are obtained from 
\begin{equation}
\delta_{\Lambda} = \frac{\partial U_\Lambda(\rho)}{\partial\rho} \biggl|_{\rho=\rho_{0,\Lambda}} , 
\quad 
\lambda_{\Lambda} = \frac{\partial^2 U_\Lambda(\rho)}{\partial\rho^2} \biggl|_{\rho=\rho_{0,\Lambda}} ,
\end{equation}
whereas 
\begin{equation}
\begin{split}
Z_{C,\Lambda} &= \lim_{p\to 0} \frac{\partial}{\partial \wm} \Gamma^{(2)}_{\Lambda,12}(p) , \\ 
V_{A,\Lambda} &= \lim_{p\to 0} \frac{\partial}{\partial \wm^2} \Gamma^{(2)}_{\Lambda,22}(p) , \\
Z_{A,\Lambda} &= \lim_{p\to 0} \frac{\partial}{\partial \p^2} \Gamma^{(2)}_{\Lambda,22}(p)  .
\end{split}
\end{equation}
$U_\Lambda$ is the mean-field effective potential and 
\begin{align}
\Gamma^{(2)}_{\Lambda,ij}(p) &= \frac{\delta^2 \Gamma_\Lambda[\Delta^*,\Delta]}{\delta\Delta_i(-p)\delta\Delta_j(p)} \biggl|_{\Delta(x)=\Delta_\Lambda}   \nonumber \\ 
&= \frac{\delta_{i,j}}{g}  + \frac{(-1)^{i+j}}{2} \int_q \tr [ \calG_\Lambda(q) \tau^i \calG_\Lambda(q+p) \tau^j ]
\label{Gam2} 
\end{align}
the mean-field two-point vertex in a uniform and time-independent field $\Delta$ (see appendix). 
We use the notation $p=(\p,i\wm)$ and $q=(\q,i\wn)$ with $\wm$ ($\wn$) a bosonic (fermionic) Matsubara frequency. $\calG_\Lambda^{-1}(q)=\calG_0^{-1}(q)+\Delta_\Lambda\tau^1$ is the mean-field fermion propagator and $(\tau^1,\tau^2,\tau^3)$ stands for the Pauli matrices. $\Delta_1$ and $\Delta_2$ refer to the real and imaginary parts of the pairing field $\Delta=(\Delta_1+i\Delta_2)/\sqrt{2}$. With no loss of generality, we have assumed $\Delta_\Lambda=\rho_{0,\Lambda}^{1/2}$ to be real. 

The bosonic effective action~(\ref{action2}) has been studied in the context of superfluidity in Bose gases and we refer to previous publications for details about the solution of the flow equation~(\ref{floweq})~\cite{Wetterich08,Floerchinger08,Dupuis09b,Sinner10}. We have used an exponential cutoff function,
\begin{equation}
\begin{gathered} 
R_k(p) = Z_{A,k} Y r(Y) , \quad r(Y) = (e^Y-1)^{-1} , \\ 
Y = \frac{\p^2}{k^2} + \frac{\wm^2}{k^2c_\Lambda^2} , 
\end{gathered} 
\label{Rk}
\end{equation}
acting both on momenta and frequencies~\cite{Dupuis09b}.

\section{Critical region of the unitary Fermi gas} 
\label{sec_unitary}

To study the unitary limit, we compute both the effective potential $U_\Lambda$ and the two-point vertex $\Gamma^{(2)}_\Lambda$ in the limit $\Lambda_F\to\infty$ and $g\to 0$ in such a way that $1/a=0$ (see appendix). We then fix the chemical potential $\mu$ to an arbitrary value and determine the superfluid transition temperature $T_c$ from the vanishing of $\rho_{0,k}$ for $k\to 0$. Finally we fine tune the value of the ultraviolet cutoff $\Lambda$ so as to obtain $T_c=0.15E_F$ where $n=-dU_{k=0}(\rho_{0,k=0})/d\mu$ is the fermion density and $E_F=k_F^2/2m=(3\pi^2n)^{2/3}/2m$. We find $\Lambda\simeq 0.97 k_F$ to be of the order of $\Delta_\Lambda(T=0)/c_\Lambda=0.77k_F$~\cite{note5}. When we compute physical quantities (e.g., the correlation length) away from the transition temperature, we adjust the value of the chemical potential so as to maintain the fermion density $n$ constant.

\begin{figure}
\centerline{\includegraphics[width=6cm]{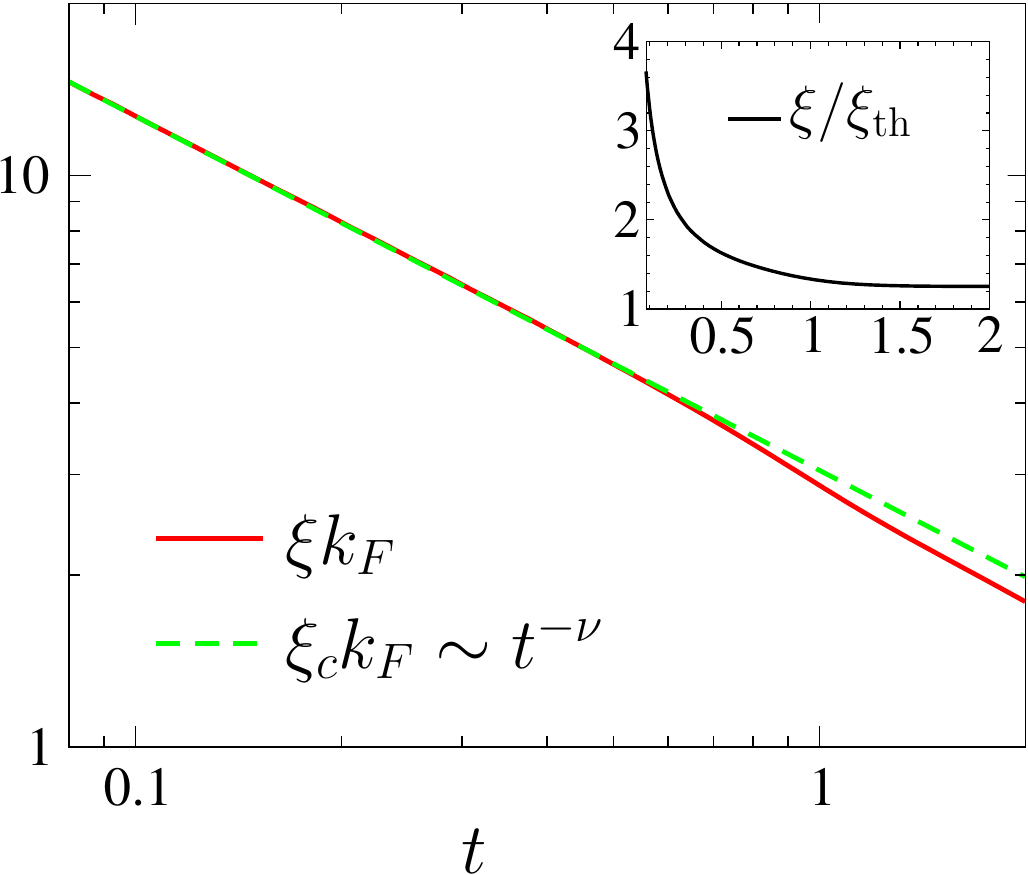}} 
\caption{(Color online) Superfluid correlation length $\xi$ vs $t=(T-T_c)/T_c$ in the normal phase ($T>T_c$). The (green) dashed line shows the critical behavior $\xi\sim t^{-\nu}$ when $t\to 0^+$. 
(Inset) Ratio between the correlation length $\xi$ and the fermion de Broglie thermal wavelength $\xi_{\rm th}=v_F/\pi T$.} 
\label{fig_xi} 
\vspace{0.25cm}
\centerline{\includegraphics[width=6cm]{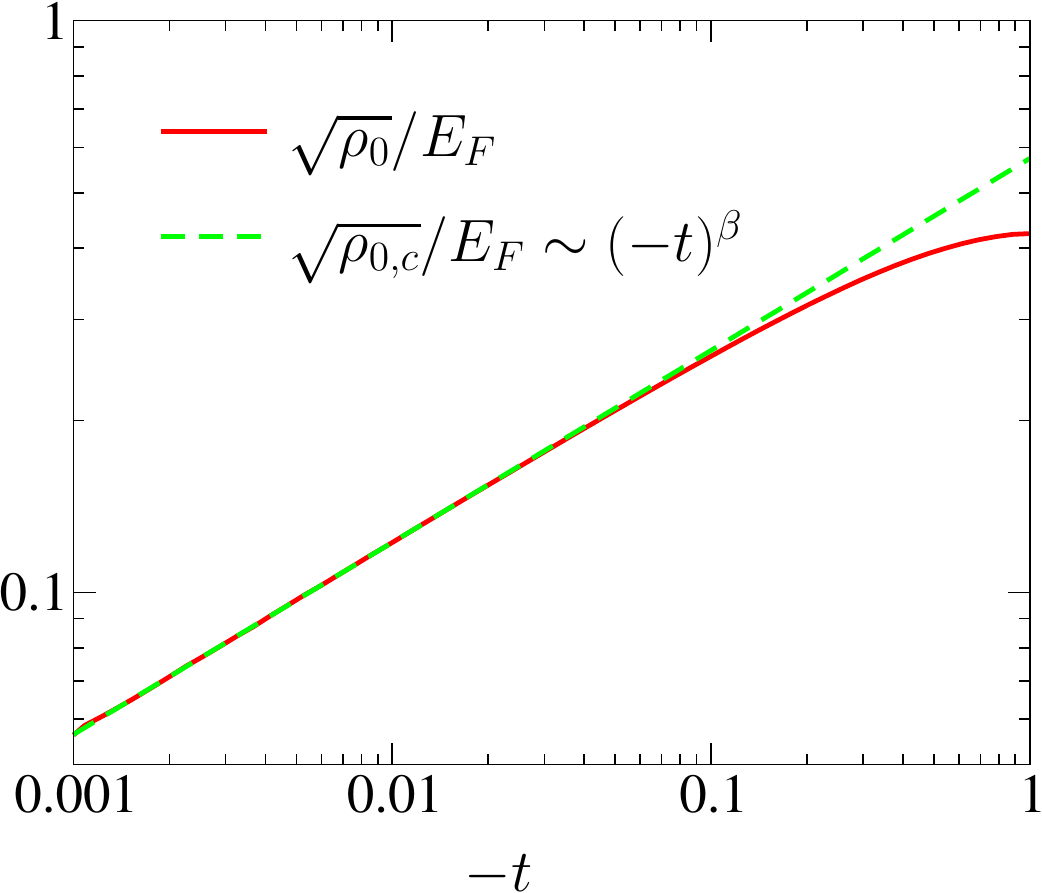}} 
\caption{(Color online) Order parameter $\sqrt{\rho_0}$ vs $t=(T-T_c)/T_c$ in the superfluid phase ($T<T_c$). The (green) dashed line shows the critical behavior $\sqrt{\rho_0}\sim (-t)^{\beta}$ when $t\to 0^-$.}
\label{fig_rho}
\end{figure}

Figure~\ref{fig_xi} shows the superfluid correlation length 
\begin{equation}
\xi = \left( \frac{Z_{A,k=0}}{\delta_{k=0}} \right)^{1/2} 
\end{equation}
in the normal phase. Near $T_c$, $\xi\sim (T-T_c)^{-\nu}$ diverges with the critical exponent $\nu\simeq 0.62$, which is the known result for the three-dimensional classical O(2) model with the approximations used here to solve the RG equation~(\ref{floweq}). Defining the Ginzburg temperature $T_G^+$ by the criterion
\begin{equation} 
\frac{\ln(\xi k_F)-\ln(\xi_c k_F)}{\ln(\xi_c k_F)} = 0.02, 
\label{TGcrit} 
\end{equation}
where $\xi_c\sim t^{-\nu}$ corresponds to the asymptotic behavior of the correlation length for $t\to 0$,
we find that the critical region $[T_c,T_G^+]$ extends up to the Ginzburg temperature $T_G^+\simeq 1.75T_c\simeq 0.26E_F$. The Ginzburg temperature is of course sensitive to the precise criterion that we use. For example, one finds $T_G^+\simeq 1.66 T_c$ (and $T_G^-\simeq 0.94 T_c$) with $0.01$ instead of $0.02$ in~(\ref{TGcrit}), and $T_G^+\simeq 1.97 T_c$ (and $T_G^-\simeq 0.79 T_c$) with $0.05$.

Figure~\ref{fig_xi} (inset) also shows the ratio between the correlation length $\xi$ and the fermion de Broglie thermal length $\xi_{\rm th}=v_F/\pi T$ ($v_F=k_F/m$). $\xi/\xi_{\rm th}$ is of order one for $T\gtrsim T_G^+$ but rapidly increases for temperatures below $T_G^+$. 

We show in Fig.~\ref{fig_rho} the order parameter $\sqrt{\rho_0}\equiv\sqrt{\rho_{0,k=0}}$ in the superfluid phase. Near $T_c$, $\sqrt{\rho_0}\sim (T_c-T)^{\beta}$ vanishes with the exponent $\beta\simeq 0.33$. The critical behavior extends from the transition temperature $T_c=0.15E_F$ down to the Ginzburg temperature $T_G^-\simeq 0.90 T_c\simeq 0.14 E_F$. The latter is obtained from the criterion~(\ref{TGcrit}) with $\xi k_F$ replaced by $\sqrt{\rho_0}/E_F$ and $\sqrt{\rho_{0,c}}\sim (-t)^\beta$.  

We conclude that the critical region $[T_G^-,T_G^+]$ is sizable and strongly asymmetric: $(T_G^+-T_c)/(T_c-T_G^-)\sim 8$. This asymmetry is partially explained by the fact that we vary the temperature at fixed density while the chemical potential $\mu$ strongly varies with $T$ in the normal phase. At fixed chemical potential we find a smaller asymmetry: $(T_G^+-T_c)/(T_c-T_G^-)\simeq 3$.   

To clarify the origin of this asymmetry we have studied the classical O(2) model ($\varphi^4$ theory for a two-component field) using the NPRG. We have found that the asymmetry is small when the interaction is weak but can be large when the interaction is strong. This shows that the critical regime is generically asymmetric when the interaction is of order one, which is the case in the unitary Fermi gas. However, in the O(2) model the critical regime is larger in the low-temperature phase, i.e., $T_G^- > T_G^+$, while the reverse is true in the unitary Fermi gas. A crucial difference between the O(2) model and our theory of the unitary Fermi gas is that in the latter case, the initial condition of the RG approach is given by the mean-field BCS theory and is temperature dependent. By contrast, the only temperature dependence in the O(2) model comes from the coefficient of the quadratic ($\varphi^2$) term. This, we believe, might explain the difference between the O(2) model and the unitary Fermi 
gas but we have not been able to find a simple physical explanation.

\section{Critical region in the BCS-BEC crossover} 

{\it A priori} it should be possible to follow a similar procedure to compute the size of the critical regime in the whole BCS-BEC crossover. While this is indeed the case on the BCS side (see below), in the BEC limit we encounter the following difficulty. When $1/k_Fa\gg 1$, superfluidity is due to the BEC of pointlike composite bosons. The initial effective action then corresponds to a bosonic action with a local interaction. To leading order in $n^{1/3}a$, the transition temperature is given by the BEC temperature of bosons with mass $m_B=2m$ and density $n_B=n/2$. 
As is well known in the theory of the dilute Bose gas, the BEC temperature is not correctly computed if one imposes a cutoff on frequencies. Technically this is due to the propagator of the bosonic field (i.e., $\Delta$) vanishing as $1/i\w_m$ in the large frequency limit. The calculation of the particle density then requires one to sum over all Matsubara frequencies (with an appropriate convergence factor). 
We expect that a cutoff acting only on momenta would allow us to explore the BEC limit but one would then have to go beyond the derivative expansion (for the frequency dependence), a very difficult task in practice.

To estimate the critical regime in the BCS-BEC crossover and in particular in the BEC limit, it is possible to use the Ginzburg criterion, with the additional constraint that the NPRG results should be reproduced in the unitary limit. Let us consider the classical limit 
\begin{equation}
\Gamma_{\rm cl}[\Delta^*,\Delta] = \beta \int d^3r \bigl\lbrace Z_{A} |\nablabf\Delta|^2 + U(\rho) \bigr\rbrace 
\end{equation}
of the effective action $\Gamma_{k=\Lambda}$ obtained by considering only fluctuation modes with vanishing Matsubara frequencies~\cite{note1}. We can then define two characteristic lengths. The first one is the $T=0$ healing length $\xi_h=(Z_A/\lambda\rho_0)^{1/2}$ ; the second one is the Ginzburg length $\xi_G=Z_A^2/\lambda T_c$~\cite{note2}. The size of the critical region is determined by the Ginzburg criterion 
\begin{equation}
\frac{|T^\pm_G-T_c|}{T_c} = t_G^\pm = \alpha_\pm \frac{\xi^2_h}{\xi^2_G} , 
\label{gcrit} 
\end{equation}
where $\alpha_\pm$ is a constant. Here we distinguish between the critical region in the normal ($t_G^+$) and superfluid ($t_G^-$) phases. $\xi_h$ and $\xi_G$ are computed vs $1/k_Fa$ using mean-field theory (see appendix) and we fix the constants $\alpha_\pm$ by requiring that for $1/a=0$ we reproduce the NPRG results obtained for the unitary gas in Sec.~\ref{sec_unitary}: $\alpha_+\simeq 4.0\times 10^{-3}$ and $\alpha_-\simeq 5.2\times 10^{-4}$. The asymmetry of the critical region is then independent of $1/k_Fa$ and equal to $\alpha_+/\alpha_-$. We therefore do not expect the Ginzburg criterion~(\ref{gcrit}) to be quantitatively valid beyond the BCS-BEC crossover region $1/k_F|a|\lesssim 1$.

\begin{figure}
\centerline{\includegraphics[width=6cm]{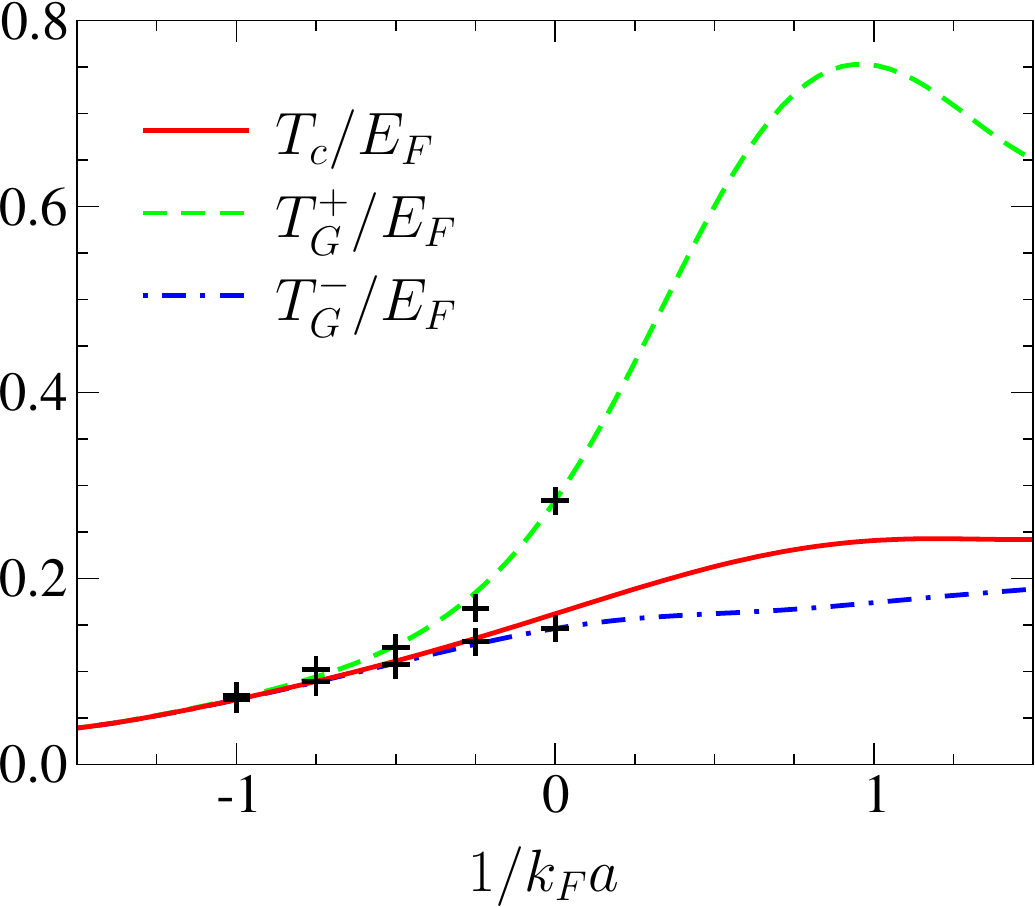}} 
\caption{(Color online) Critical region of the superfluid transition in the BCS-BEC crossover. The transition temperature $T_c$ is obtained from a Luttinger-Ward approach~\cite{Haussmann07}. The Ginzburg temperatures $T_G^+$ and $T_G^-$ (dashed and dot-dashed lines) are deduced from~(\ref{gcrit}). The black crosses on the BCS side show the results obtained directly from the NPRG approach (as in the unitary limit).} 
\label{fig_tG}
\end{figure}

Using the results of the appendix, in the BCS limit we find that $t_G^\pm\sim (T_c/E_F)^4$ is exponentially small, since $T_c\sim E_Fe^{-\pi/k_F|a|}$, so that the critical region is too small to be observed experimentally. In the BEC limit, $t_G^\pm \sim k_Fa(T_c/E_F)^2 \sim k_Fa$~\cite{note3,Taylor09} is also small although not as strongly suppressed as in the BCS limit. For $1/k_Fa\gg 1$, using $T_c\simeq 0.218 E_F$ (and the results of the appendix)~\cite{note7}, we obtain $t_G^+\simeq 7 n^{1/3}a\ll 1$ in agreement with known results of the dilute Bose gas~\cite{Giorgini96,note6}. 
By contrast, $t_G^\pm$ is of order unity at unitarity. Note that the maximum of $T_G^+$ is on the BEC side of the resonance ($1/k_Fa\simeq 1$), not far away from the maximum of $T_c$. 

The critical region in the BCS-BEC crossover is shown in Fig.~\ref{fig_tG}, where we have used the transition temperature $T_c$ computed in Ref.~\cite{Haussmann07} within a Luttinger-Ward  approach. On the BCS side $1/k_Fa<0$, we also show the critical region obtained from the NPRG approach following the procedure used in the unitary limit (the cutoff $\Lambda$ is fixed in order to reproduce the transition temperature computed in Ref.~\cite{Haussmann07}). There is a good agreement with the results obtained from the Ginzburg criterion. By contrast, on the BEC side $1/k_Fa>0$, the NPRG approach yields a much narrower critical region than what is predicted by the Ginzburg criterion. We ascribe this disagreement to the failure of our NPRG approach (with a frequency cutoff) to determine correctly the particle density when there are composite bosons (bound pairs of fermions); see the discussion at the beginning of this section. 

The overall picture of the critical region is in qualitative agreement with the results of Ref.~\cite{Taylor09}. Evidence for a large critical region close to unitarity was also found in Ref.~\cite{Strinati02}, where a diagrammatic analysis showed that diagrams that are subleading for $T\gg T_c$ exceed leading ones in a wide region, of order $T_c$, above the transition temperature.

\section{Conclusion}

In summary, we have determined the width of the critical region in a Fermi gas through the BCS-BEC crossover. To circumvent the breakdown of perturbation theory in the critical region, we have used a NPRG approach. Close to unitarity, we find that the critical region is very wide (of the order of the transition temperature $T_c$) and asymmetric, being larger in the normal phase than in the superfluid phase. The critical region is also more pronounced on the BEC side than on the BCS side where it becomes unobservable. 
An interesting and open issue relates to the interplay between critical fluctuations, Fermi-liquid behavior, and the (much debated) possible existence of a pseudogap~\cite{[{See, for instance, }]Sun15}. 

In Ref.~\cite{Taylor09}, it was pointed out that the critical region is strongly suppressed in trapped gases due to their inhomogeneity, although critical behavior could be observed in the compressibility by measuring the density profile. Nonetheless, the possibility to load gases in the three-dimensional quasiuniform potential of an optical box trap opens up new prospects to observe criticality in trapped gases~\cite{Gaunt13}.

\acknowledgments
We would like to thank A.-M. Tremblay for useful discussions on the pseudogap issue in fermion systems.

\appendix

\section*{Appendix: Effective potential and two-point vertex in the mean-field approximation}

At the initial stage of the RG procedure, the effective action is given by mean-field theory: $\Gamma_\Lambda[\Delta^*,\Delta]=S[\Delta^*,\Delta]$, where $S$ is given by~(\ref{action3}). An elementary calculation then gives
\begin{equation}
U_\Lambda(\rho,\mu) = \frac{1}{g} + \int_\q \llbrace \xiq  - 2 T \ln[ 2 \cosh(\beta\Eq/2)] \rrbrace 
\label{app1}
\end{equation}
and 
\begin{align} 
\Gamma^{(2)}_{\Lambda,\rm 11}(p;\rho) ={}& \frac{1}{g} + \half \int_q \lbrace 2 F(q)F(p+q) \nonumber \\
& - G(q)[G(p-q)+G(-p-q)] \rbrace , \nonumber\\ 
\Gamma^{(2)}_{\Lambda,\rm 22}(p;\rho) ={}& \frac{1}{g} - \half \int_q \lbrace 2 F(q)F(p+q) \label{app2} \\
& + G(q)[G(p-q)+G(-p-q)] \rbrace , \nonumber\\
\Gamma^{(2)}_{\Lambda,\rm 12}(p;\rho) ={}& -\frac{i}{2} \int_q G(q)[G(p-q)-G(-p-q)] , \nonumber
\end{align}
in a constant field $\Delta=\sqrt{\rho}$ (i.e., $\Delta_1=\sqrt{2\rho}$ and $\Delta_2=0$), with 
\begin{align}
G(q) = - \frac{i\wn+\xiq}{\wn^2+\Eq^2} , \qquad F(q) =  \frac{\sqrt{\rho}}{\wn^2+\Eq^2} , \nonumber\\
\Eq=\sqrt{\xiq^2+\rho} , \qquad \xiq = \eps_\q - \mu . 
\end{align} 
We use the notation
\begin{equation}
\int_q \equiv T \sum_{\wn} \int_\q \equiv T \sum_{\wn} \int \frac{d^3q}{(2\pi)^3} . 
\end{equation}
The fermionic Matsubara sums in~(\ref{app2}) can be done analytically. The interaction constant $g$ can be eliminated using~(\ref{adef}). We then take the limit $\Lambda_F\to\infty$ in the momentum integrals with $a$ fixed. 

\subsection*{Derivative expansion}

Equations~(\ref{app1}) and (\ref{app2}) yield 
\begin{widetext}
\begin{equation}
\begin{split}
\lamb_\Lambda ={}& \quarter \int_\q \left[ \frac{1}{\Eq^3} \tanh(\beta\Eq/2) - \frac{\beta}{2\Eq^2} \cosh^{-2}(\beta\Eq/2) \right], \\ 
Z_{A,\Lambda} ={}& \frac{1}{8m} \int_\q \left( \xiq + \frac{2\q^2}{dm} \right) \left[ \frac{1}{\Eq^3} \tanh(\beta\Eq/2) - \frac{\beta}{2\Eq^2} \cosh^{-2}(\beta\Eq/2) \right] \\ & 
- \frac{1}{8m^2d} \int_\q \q^2\xiq^2 \left[ \frac{3}{\Eq^5} \tanh(\beta\Eq/2) - \frac{3\beta}{2\Eq^4} \cosh^{-2}(\beta\Eq/2)- \frac{\beta^2}{2\Eq^3} \frac{\tanh(\beta\Eq/2)}{\cosh^2(\beta\Eq/2)} \right] , \\
Z_{C,\Lambda} ={}& \quarter \int_\q \xiq \left[ \frac{1}{\Eq^3} \tanh(\beta\Eq/2) - \frac{\beta}{2\Eq^2} \cosh^{-2}(\beta\Eq/2) \right], \\ 
V_{A,\Lambda} ={}& \frac{1}{8} \int_\q \left[ \frac{1}{\Eq^3} \tanh(\beta\Eq/2) - \frac{\beta}{2\Eq^2} \cosh^{-2}(\beta\Eq/2) - \frac{\beta^2}{2\Eq} \frac{\tanh(\beta\Eq/2)}{\cosh^2(\beta\Eq/2)} \right] ,
\end{split}
\end{equation}
\end{widetext}
where $d=3$ and the right-hand side is evaluated at the minimum of the effective potential defined by $U'_\Lambda(\rho_{0,\Lambda})=0$. 

In the $T=0$ BCS limit, we obtain 
\begin{equation}
U_\Lambda(\rho) \simeq -\Theta(\mu) \frac{2}{15\pi^2} (2m)^{3/2} \mu^{5/2} ,
\end{equation}
and 
\begin{equation}
\begin{gathered}
\sqrt{\rho_{0,\Lambda}} = 8e^{-2} \eps_F e^{-\pi/2k_F|a|} , \quad \lamb_\Lambda = \frac{mk_F}{4\pi^2\rho_{0,\Lambda}} , \\ 
Z_{A,\Lambda} = \frac{\bar n}{8m\rho_{0,\Lambda}} , \quad
Z_{C,\Lambda} \simeq 0 , \quad 
V_{A,\Lambda} = \frac{mk_F}{8\pi^2\rho_{0,\Lambda}} , 
\end{gathered}
\end{equation} 
where $\bar n=-d U_\Lambda/d\mu=k_F^3/3\pi^2$, $k_F=\sqrt{2m\eps_F}$, and $\eps_F=\mu$. 

In the $T=0$ BEC limit, we find 
\begin{equation} 
U_\Lambda(\rho) = - \frac{m\rho}{4\pi a} + \frac{\rho(2m)^{3/2}|\mu|^{1/2}}{8\pi} + \frac{\rho^2(2m)^{3/2}|\mu|^{-3/2}}{256\pi} 
\end{equation}
and
\begin{equation}
\begin{gathered}
\rho_{0,\Lambda} = \frac{4\pi \bar n}{m^2a} = \frac{4}{ma^2} \delta\mu, \quad \lamb_\Lambda = \frac{m^3 a^3}{16\pi} , \\ 
Z_{A,\Lambda} = \frac{ma}{32\pi} , \quad Z_{C,\Lambda} = \frac{m^2a}{8\pi} , \quad V_{A,\Lambda} = \frac{m^3a^3}{32\pi} ,
\end{gathered}
\end{equation}
where $\mu=-1/2ma^2+\delta\mu$ ($0\leq \delta\mu \ll 1/2ma^2$). 



\begin{thebibliography}{43}%
\makeatletter
\providecommand \@ifxundefined [1]{%
 \@ifx{#1\undefined}
}%
\providecommand \@ifnum [1]{%
 \ifnum #1\expandafter \@firstoftwo
 \else \expandafter \@secondoftwo
 \fi
}%
\providecommand \@ifx [1]{%
 \ifx #1\expandafter \@firstoftwo
 \else \expandafter \@secondoftwo
 \fi
}%
\providecommand \natexlab [1]{#1}%
\providecommand \enquote  [1]{``#1''}%
\providecommand \bibnamefont  [1]{#1}%
\providecommand \bibfnamefont [1]{#1}%
\providecommand \citenamefont [1]{#1}%
\providecommand \href@noop [0]{\@secondoftwo}%
\providecommand \href [0]{\begingroup \@sanitize@url \@href}%
\providecommand \@href[1]{\@@startlink{#1}\@@href}%
\providecommand \@@href[1]{\endgroup#1\@@endlink}%
\providecommand \@sanitize@url [0]{\catcode `\\12\catcode `\$12\catcode
  `\&12\catcode `\#12\catcode `\^12\catcode `\_12\catcode `\%12\relax}%
\providecommand \@@startlink[1]{}%
\providecommand \@@endlink[0]{}%
\providecommand \url  [0]{\begingroup\@sanitize@url \@url }%
\providecommand \@url [1]{\endgroup\@href {#1}{\urlprefix }}%
\providecommand \urlprefix  [0]{URL }%
\providecommand \Eprint [0]{\href }%
\providecommand \doibase [0]{http://dx.doi.org/}%
\providecommand \selectlanguage [0]{\@gobble}%
\providecommand \bibinfo  [0]{\@secondoftwo}%
\providecommand \bibfield  [0]{\@secondoftwo}%
\providecommand \translation [1]{[#1]}%
\providecommand \BibitemOpen [0]{}%
\providecommand \bibitemStop [0]{}%
\providecommand \bibitemNoStop [0]{.\EOS\space}%
\providecommand \EOS [0]{\spacefactor3000\relax}%
\providecommand \BibitemShut  [1]{\csname bibitem#1\endcsname}%
\let\auto@bib@innerbib\@empty
\bibitem [{\citenamefont {Bardeen}\ \emph {et~al.}(1957)\citenamefont
  {Bardeen}, \citenamefont {Cooper},\ and\ \citenamefont
  {Schrieffer}}]{Bardeen57}%
  \BibitemOpen
  \bibfield  {author} {\bibinfo {author} {\bibfnamefont {J.}~\bibnamefont
  {Bardeen}}, \bibinfo {author} {\bibfnamefont {L.~N.}\ \bibnamefont {Cooper}},
  \ and\ \bibinfo {author} {\bibfnamefont {J.~R.}\ \bibnamefont {Schrieffer}},\
  }\href {\doibase 10.1103/PhysRev.108.1175} {\bibfield  {journal} {\bibinfo
  {journal} {Phys. Rev.}\ }\textbf {\bibinfo {volume} {108}},\ \bibinfo {pages}
  {1175} (\bibinfo {year} {1957})}\BibitemShut {NoStop}%
\bibitem [{\citenamefont {Bogoliubov}(1947)}]{Bogoliubov47}%
  \BibitemOpen
  \bibfield  {author} {\bibinfo {author} {\bibfnamefont {N.~N.}\ \bibnamefont
  {Bogoliubov}},\ }\href@noop {} {\bibfield  {journal} {\bibinfo  {journal} {J.
  Phys. (USSR)}\ }\textbf {\bibinfo {volume} {11}},\ \bibinfo {pages} {23}
  (\bibinfo {year} {1947})}\BibitemShut {NoStop}%
\bibitem [{\citenamefont {Giorgini}\ \emph {et~al.}(2008)\citenamefont
  {Giorgini}, \citenamefont {Pitaevskii},\ and\ \citenamefont
  {Stringari}}]{Giorgini08}%
  \BibitemOpen
  \bibfield  {author} {\bibinfo {author} {\bibfnamefont {S.}~\bibnamefont
  {Giorgini}}, \bibinfo {author} {\bibfnamefont {L.~P.}\ \bibnamefont
  {Pitaevskii}}, \ and\ \bibinfo {author} {\bibfnamefont {S.}~\bibnamefont
  {Stringari}},\ }\href {\doibase 10.1103/RevModPhys.80.1215} {\bibfield
  {journal} {\bibinfo  {journal} {Rev. Mod. Phys.}\ }\textbf {\bibinfo {volume}
  {80}},\ \bibinfo {pages} {1215} (\bibinfo {year} {2008})}\BibitemShut
  {NoStop}%
\bibitem [{\citenamefont {Chen}\ \emph {et~al.}(2005)\citenamefont {Chen},
  \citenamefont {Stajic}, \citenamefont {Tan},\ and\ \citenamefont
  {Levin}}]{Chen05}%
  \BibitemOpen
  \bibfield  {author} {\bibinfo {author} {\bibfnamefont {Q.}~\bibnamefont
  {Chen}}, \bibinfo {author} {\bibfnamefont {J.}~\bibnamefont {Stajic}},
  \bibinfo {author} {\bibfnamefont {S.}~\bibnamefont {Tan}}, \ and\ \bibinfo
  {author} {\bibfnamefont {K.}~\bibnamefont {Levin}},\ }\href {\doibase
  http://dx.doi.org/10.1016/j.physrep.2005.02.005} {\bibfield  {journal}
  {\bibinfo  {journal} {Physics Reports}\ }\textbf {\bibinfo {volume} {412}},\
  \bibinfo {pages} {1 } (\bibinfo {year} {2005})}\BibitemShut {NoStop}%
\bibitem [{\citenamefont {Bloch}\ \emph {et~al.}(2008)\citenamefont {Bloch},
  \citenamefont {Dalibard},\ and\ \citenamefont {Zwerger}}]{Bloch08}%
  \BibitemOpen
  \bibfield  {author} {\bibinfo {author} {\bibfnamefont {I.}~\bibnamefont
  {Bloch}}, \bibinfo {author} {\bibfnamefont {J.}~\bibnamefont {Dalibard}}, \
  and\ \bibinfo {author} {\bibfnamefont {W.}~\bibnamefont {Zwerger}},\ }\href
  {\doibase 10.1103/RevModPhys.80.885} {\bibfield  {journal} {\bibinfo
  {journal} {Rev. Mod. Phys.}\ }\textbf {\bibinfo {volume} {80}},\ \bibinfo
  {pages} {885} (\bibinfo {year} {2008})}\BibitemShut {NoStop}%
\bibitem [{\citenamefont {Zwerger}(2012)}]{Zwerger12}%
  \BibitemOpen
  \bibinfo {editor} {\bibfnamefont {W.}~\bibnamefont {Zwerger}},\ editor,\
  \href@noop {} {\emph {\bibinfo {title} {The BCS-BEC Crossover and the
  Unitarity Fermi Gas}}}\ (\bibinfo  {publisher} {Lecture Notes in Physics,
  Springer, New York},\ \bibinfo {year} {2012})\BibitemShut {NoStop}%
\bibitem [{\citenamefont {Randeria}\ and\ \citenamefont
  {Taylor}(2014)}]{Randeria14}%
  \BibitemOpen
  \bibfield  {author} {\bibinfo {author} {\bibfnamefont {M.}~\bibnamefont
  {Randeria}}\ and\ \bibinfo {author} {\bibfnamefont {E.}~\bibnamefont
  {Taylor}},\ }\href {\doibase 10.1146/annurev-conmatphys-031113-133829}
  {\bibfield  {journal} {\bibinfo  {journal} {Annual Review of Condensed Matter
  Physics}\ }\textbf {\bibinfo {volume} {5}},\ \bibinfo {pages} {209} (\bibinfo
  {year} {2014})}\BibitemShut {NoStop}%
\bibitem [{not({\natexlab{a}})}]{note8}%
  \BibitemOpen
  \href@noop {} {} \bibinfo {note} {The highest $T_c$ is
  actually reached slightly on the BEC side of the Feshbach
  resonance.}\BibitemShut {Stop}%
\bibitem [{\citenamefont {Stajic}\ \emph {et~al.}(2004)\citenamefont {Stajic},
  \citenamefont {Milstein}, \citenamefont {Chen}, \citenamefont {Chiofalo},
  \citenamefont {Holland},\ and\ \citenamefont {Levin}}]{Stajic04}%
  \BibitemOpen
  \bibfield  {author} {\bibinfo {author} {\bibfnamefont {J.}~\bibnamefont
  {Stajic}}, \bibinfo {author} {\bibfnamefont {J.~N.}\ \bibnamefont
  {Milstein}}, \bibinfo {author} {\bibfnamefont {Q.}~\bibnamefont {Chen}},
  \bibinfo {author} {\bibfnamefont {M.~L.}\ \bibnamefont {Chiofalo}}, \bibinfo
  {author} {\bibfnamefont {M.~J.}\ \bibnamefont {Holland}}, \ and\ \bibinfo
  {author} {\bibfnamefont {K.}~\bibnamefont {Levin}},\ }\href {\doibase
  10.1103/PhysRevA.69.063610} {\bibfield  {journal} {\bibinfo  {journal} {Phys.
  Rev. A}\ }\textbf {\bibinfo {volume} {69}},\ \bibinfo {pages} {063610}
  (\bibinfo {year} {2004})}\BibitemShut {NoStop}%
\bibitem [{\citenamefont {Perali}\ \emph {et~al.}(2004)\citenamefont {Perali},
  \citenamefont {Pieri}, \citenamefont {Pisani},\ and\ \citenamefont
  {Strinati}}]{Perali04}%
  \BibitemOpen
  \bibfield  {author} {\bibinfo {author} {\bibfnamefont {A.}~\bibnamefont
  {Perali}}, \bibinfo {author} {\bibfnamefont {P.}~\bibnamefont {Pieri}},
  \bibinfo {author} {\bibfnamefont {L.}~\bibnamefont {Pisani}}, \ and\ \bibinfo
  {author} {\bibfnamefont {G.~C.}\ \bibnamefont {Strinati}},\ }\href {\doibase
  10.1103/PhysRevLett.92.220404} {\bibfield  {journal} {\bibinfo  {journal}
  {Phys. Rev. Lett.}\ }\textbf {\bibinfo {volume} {92}},\ \bibinfo {pages}
  {220404} (\bibinfo {year} {2004})}\BibitemShut {NoStop}%
\bibitem [{\citenamefont {Perali}\ \emph {et~al.}(2011)\citenamefont {Perali},
  \citenamefont {Palestini}, \citenamefont {Pieri}, \citenamefont {Strinati},
  \citenamefont {Stewart}, \citenamefont {Gaebler}, \citenamefont {Drake},\
  and\ \citenamefont {Jin}}]{Perali11}%
  \BibitemOpen
  \bibfield  {author} {\bibinfo {author} {\bibfnamefont {A.}~\bibnamefont
  {Perali}}, \bibinfo {author} {\bibfnamefont {F.}~\bibnamefont {Palestini}},
  \bibinfo {author} {\bibfnamefont {P.}~\bibnamefont {Pieri}}, \bibinfo
  {author} {\bibfnamefont {G.~C.}\ \bibnamefont {Strinati}}, \bibinfo {author}
  {\bibfnamefont {J.~T.}\ \bibnamefont {Stewart}}, \bibinfo {author}
  {\bibfnamefont {J.~P.}\ \bibnamefont {Gaebler}}, \bibinfo {author}
  {\bibfnamefont {T.~E.}\ \bibnamefont {Drake}}, \ and\ \bibinfo {author}
  {\bibfnamefont {D.~S.}\ \bibnamefont {Jin}},\ }\href {\doibase
  10.1103/PhysRevLett.106.060402} {\bibfield  {journal} {\bibinfo  {journal}
  {Phys. Rev. Lett.}\ }\textbf {\bibinfo {volume} {106}},\ \bibinfo {pages}
  {060402} (\bibinfo {year} {2011})}\BibitemShut {NoStop}%
\bibitem [{\citenamefont {Nascimb\`ene}\ \emph {et~al.}(2011)\citenamefont
  {Nascimb\`ene}, \citenamefont {Navon}, \citenamefont {Pilati}, \citenamefont
  {Chevy}, \citenamefont {Giorgini}, \citenamefont {Georges},\ and\
  \citenamefont {Salomon}}]{Nascimbene11}%
  \BibitemOpen
  \bibfield  {author} {\bibinfo {author} {\bibfnamefont {S.}~\bibnamefont
  {Nascimb\`ene}}, \bibinfo {author} {\bibfnamefont {N.}~\bibnamefont {Navon}},
  \bibinfo {author} {\bibfnamefont {S.}~\bibnamefont {Pilati}}, \bibinfo
  {author} {\bibfnamefont {F.}~\bibnamefont {Chevy}}, \bibinfo {author}
  {\bibfnamefont {S.}~\bibnamefont {Giorgini}}, \bibinfo {author}
  {\bibfnamefont {A.}~\bibnamefont {Georges}}, \ and\ \bibinfo {author}
  {\bibfnamefont {C.}~\bibnamefont {Salomon}},\ }\href {\doibase
  10.1103/PhysRevLett.106.215303} {\bibfield  {journal} {\bibinfo  {journal}
  {Phys. Rev. Lett.}\ }\textbf {\bibinfo {volume} {106}},\ \bibinfo {pages}
  {215303} (\bibinfo {year} {2011})}\BibitemShut {NoStop}%
\bibitem [{\citenamefont {Gaebler}\ \emph {et~al.}(2010)\citenamefont
  {Gaebler}, \citenamefont {Stewart}, \citenamefont {Drake}, \citenamefont
  {Jin}, \citenamefont {Perali}, \citenamefont {Pieri},\ and\ \citenamefont
  {Strinati}}]{Gaebler10}%
  \BibitemOpen
  \bibfield  {author} {\bibinfo {author} {\bibfnamefont {J.~P.}\ \bibnamefont
  {Gaebler}}, \bibinfo {author} {\bibfnamefont {J.~T.}\ \bibnamefont
  {Stewart}}, \bibinfo {author} {\bibfnamefont {T.~E.}\ \bibnamefont {Drake}},
  \bibinfo {author} {\bibfnamefont {D.~S.}\ \bibnamefont {Jin}}, \bibinfo
  {author} {\bibfnamefont {A.}~\bibnamefont {Perali}}, \bibinfo {author}
  {\bibfnamefont {P.}~\bibnamefont {Pieri}}, \ and\ \bibinfo {author}
  {\bibfnamefont {G.~C.}\ \bibnamefont {Strinati}},\ }\href {\doibase
  doi:10.1038/nphys1709} {\bibfield  {journal} {\bibinfo  {journal} {Nature
  Phys.}\ }\textbf {\bibinfo {volume} {6}},\ \bibinfo {pages} {569} (\bibinfo
  {year} {2010})}\BibitemShut {NoStop}%
\bibitem [{\citenamefont {Cao}\ \emph {et~al.}(2011)\citenamefont {Cao},
  \citenamefont {Elliott}, \citenamefont {Joseph}, \citenamefont {Wu},
  \citenamefont {Petricka}, \citenamefont {Schaefer},\ and\ \citenamefont
  {E.Thomas}}]{Cao11}%
  \BibitemOpen
  \bibfield  {author} {\bibinfo {author} {\bibfnamefont {C.}~\bibnamefont
  {Cao}}, \bibinfo {author} {\bibfnamefont {E.}~\bibnamefont {Elliott}},
  \bibinfo {author} {\bibfnamefont {J.}~\bibnamefont {Joseph}}, \bibinfo
  {author} {\bibfnamefont {H.}~\bibnamefont {Wu}}, \bibinfo {author}
  {\bibfnamefont {J.}~\bibnamefont {Petricka}}, \bibinfo {author}
  {\bibfnamefont {T.}~\bibnamefont {Schaefer}}, \ and\ \bibinfo {author}
  {\bibfnamefont {J.}~\bibnamefont {E.Thomas}},\ }\href {\doibase
  10.1126/science.1195219} {\bibfield  {journal} {\bibinfo  {journal}
  {Science}\ }\textbf {\bibinfo {volume} {331}},\ \bibinfo {pages} {58}
  (\bibinfo {year} {2011})}\BibitemShut {NoStop}%
\bibitem [{\citenamefont {Nascimb\`ene}\ \emph {et~al.}(2010)\citenamefont
  {Nascimb\`ene}, \citenamefont {Navon}, \citenamefont {Jiang}, \citenamefont
  {Chevy},\ and\ \citenamefont {Salomon}}]{Nascimbene10}%
  \BibitemOpen
  \bibfield  {author} {\bibinfo {author} {\bibfnamefont {S.}~\bibnamefont
  {Nascimb\`ene}}, \bibinfo {author} {\bibfnamefont {N.}~\bibnamefont {Navon}},
  \bibinfo {author} {\bibfnamefont {K.~J.}\ \bibnamefont {Jiang}}, \bibinfo
  {author} {\bibfnamefont {F.}~\bibnamefont {Chevy}}, \ and\ \bibinfo {author}
  {\bibfnamefont {C.}~\bibnamefont {Salomon}},\ }\href {\doibase
  10.1038/nature08814} {\bibfield  {journal} {\bibinfo  {journal} {Nature (London)}\
  }\textbf {\bibinfo {volume} {463}},\ \bibinfo {pages} {1057} (\bibinfo {year}
  {2010})}\BibitemShut {NoStop}%
\bibitem [{\citenamefont {Navon}\ \emph {et~al.}(2010)\citenamefont {Navon},
  \citenamefont {Nascimb\`ene}, \citenamefont {Chevy},\ and\ \citenamefont
  {Salomon}}]{Navon10}%
  \BibitemOpen
  \bibfield  {author} {\bibinfo {author} {\bibfnamefont {N.}~\bibnamefont
  {Navon}}, \bibinfo {author} {\bibfnamefont {S.}~\bibnamefont {Nascimb\`ene}},
  \bibinfo {author} {\bibfnamefont {F.}~\bibnamefont {Chevy}}, \ and\ \bibinfo
  {author} {\bibfnamefont {C.}~\bibnamefont {Salomon}},\ }\href {\doibase
  10.1126/science.1187582} {\bibfield  {journal} {\bibinfo  {journal}
  {Science}\ }\textbf {\bibinfo {volume} {328}},\ \bibinfo {pages} {729}
  (\bibinfo {year} {2010})}\BibitemShut {NoStop}%
\bibitem [{\citenamefont {Sommer}\ \emph {et~al.}(2011)\citenamefont {Sommer},
  \citenamefont {Ku}, \citenamefont {Roati},\ and\ \citenamefont
  {Zwierlein}}]{Sommer11}%
  \BibitemOpen
  \bibfield  {author} {\bibinfo {author} {\bibfnamefont {A.}~\bibnamefont
  {Sommer}}, \bibinfo {author} {\bibfnamefont {M.}~\bibnamefont {Ku}}, \bibinfo
  {author} {\bibfnamefont {G.}~\bibnamefont {Roati}}, \ and\ \bibinfo {author}
  {\bibfnamefont {M.~W.}\ \bibnamefont {Zwierlein}},\ }\href {\doibase
  10.1038/nature09989} {\bibfield  {journal} {\bibinfo  {journal} {Nature (London)}\
  }\textbf {\bibinfo {volume} {472}},\ \bibinfo {pages} {201} (\bibinfo {year}
  {2011})}\BibitemShut {NoStop}%
\bibitem [{\citenamefont {Taylor}(2009)}]{Taylor09}%
  \BibitemOpen
  \bibfield  {author} {\bibinfo {author} {\bibfnamefont {E.}~\bibnamefont
  {Taylor}},\ }\href {\doibase 10.1103/PhysRevA.80.023612} {\bibfield
  {journal} {\bibinfo  {journal} {Phys. Rev. A}\ }\textbf {\bibinfo {volume}
  {80}},\ \bibinfo {pages} {023612} (\bibinfo {year} {2009})}\BibitemShut
  {NoStop}%
\bibitem [{\citenamefont {Nozi\`eres}\ and\ \citenamefont
  {Schmitt-Rink}(1985)}]{Nozieres85}%
  \BibitemOpen
  \bibfield  {author} {\bibinfo {author} {\bibfnamefont {P.}~\bibnamefont
  {Nozi\`eres}}\ and\ \bibinfo {author} {\bibfnamefont {S.}~\bibnamefont
  {Schmitt-Rink}},\ }\href@noop {} {\bibfield  {journal} {\bibinfo  {journal}
  {J. Low Temp. Phys.}\ }\textbf {\bibinfo {volume} {59}},\ \bibinfo {pages}
  {195} (\bibinfo {year} {1985})}\BibitemShut {NoStop}%
\bibitem [{\citenamefont {S\'a~de Melo}\ \emph {et~al.}(1993)\citenamefont
  {S\'a~de Melo}, \citenamefont {Randeria},\ and\ \citenamefont
  {Engelbrecht}}]{Sademelo93}%
  \BibitemOpen
  \bibfield  {author} {\bibinfo {author} {\bibfnamefont {C.~A.~R.}\
  \bibnamefont {S\'a~de Melo}}, \bibinfo {author} {\bibfnamefont
  {M.}~\bibnamefont {Randeria}}, \ and\ \bibinfo {author} {\bibfnamefont
  {J.~R.}\ \bibnamefont {Engelbrecht}},\ }\href {\doibase
  10.1103/PhysRevLett.71.3202} {\bibfield  {journal} {\bibinfo  {journal}
  {Phys. Rev. Lett.}\ }\textbf {\bibinfo {volume} {71}},\ \bibinfo {pages}
  {3202} (\bibinfo {year} {1993})}\BibitemShut {NoStop}%
\bibitem [{\citenamefont {Burovski}\ \emph {et~al.}(2008)\citenamefont
  {Burovski}, \citenamefont {Kozik}, \citenamefont {Prokof'ev}, \citenamefont
  {Svistunov},\ and\ \citenamefont {Troyer}}]{Burovski08}%
  \BibitemOpen
  \bibfield  {author} {\bibinfo {author} {\bibfnamefont {E.}~\bibnamefont
  {Burovski}}, \bibinfo {author} {\bibfnamefont {E.}~\bibnamefont {Kozik}},
  \bibinfo {author} {\bibfnamefont {N.}~\bibnamefont {Prokof'ev}}, \bibinfo
  {author} {\bibfnamefont {B.}~\bibnamefont {Svistunov}}, \ and\ \bibinfo
  {author} {\bibfnamefont {M.}~\bibnamefont {Troyer}},\ }\href {\doibase
  10.1103/PhysRevLett.101.090402} {\bibfield  {journal} {\bibinfo  {journal}
  {Phys. Rev. Lett.}\ }\textbf {\bibinfo {volume} {101}},\ \bibinfo {pages}
  {090402} (\bibinfo {year} {2008})}\BibitemShut {NoStop}%
\bibitem [{\citenamefont {Jakubczyk}\ \emph {et~al.}(2008)\citenamefont
  {Jakubczyk}, \citenamefont {Strack}, \citenamefont {Katanin},\ and\
  \citenamefont {Metzner}}]{Jakubczyk08a}%
  \BibitemOpen
  \bibfield  {author} {\bibinfo {author} {\bibfnamefont {P.}~\bibnamefont
  {Jakubczyk}}, \bibinfo {author} {\bibfnamefont {P.}~\bibnamefont {Strack}},
  \bibinfo {author} {\bibfnamefont {A.~A.}\ \bibnamefont {Katanin}}, \ and\
  \bibinfo {author} {\bibfnamefont {W.}~\bibnamefont {Metzner}},\ }\href
  {\doibase 10.1103/PhysRevB.77.195120} {\bibfield  {journal} {\bibinfo
  {journal} {Phys. Rev. B}\ }\textbf {\bibinfo {volume} {77}},\ \bibinfo
  {pages} {195120} (\bibinfo {year} {2008})}\BibitemShut {NoStop}%
\bibitem [{\citenamefont {Strack}\ and\ \citenamefont
  {Jakubczyk}(2009)}]{Strack09}%
  \BibitemOpen
  \bibfield  {author} {\bibinfo {author} {\bibfnamefont {P.}~\bibnamefont
  {Strack}}\ and\ \bibinfo {author} {\bibfnamefont {P.}~\bibnamefont
  {Jakubczyk}},\ }\href {\doibase 10.1103/PhysRevB.80.085108} {\bibfield
  {journal} {\bibinfo  {journal} {Phys. Rev. B}\ }\textbf {\bibinfo {volume}
  {80}},\ \bibinfo {pages} {085108} (\bibinfo {year} {2009})}\BibitemShut
  {NoStop}%
\bibitem [{\citenamefont {Scherer}\ \emph {et~al.}(2011)\citenamefont
  {Scherer}, \citenamefont {Floerchinger},\ and\ \citenamefont
  {Gies}}]{Scherer10}%
  \BibitemOpen
  \bibfield  {author} {\bibinfo {author} {\bibfnamefont {M.~M.}\ \bibnamefont
  {Scherer}}, \bibinfo {author} {\bibfnamefont {S.}~\bibnamefont
  {Floerchinger}}, \ and\ \bibinfo {author} {\bibfnamefont {H.}~\bibnamefont
  {Gies}},\ }\href {\doibase 10.1098/rsta.2011.0072} {\bibfield  {journal}
  {\bibinfo  {journal} {Phil. Trans. R. Soc. A}\ }\textbf {\bibinfo {volume}
  {369}},\ \bibinfo {pages} {2779} (\bibinfo {year} {2011})}\BibitemShut
  {NoStop}%
\bibitem [{\citenamefont {Boettcher}\ \emph {et~al.}(2012)\citenamefont
  {Boettcher}, \citenamefont {Pawlowski},\ and\ \citenamefont
  {Diehl}}]{Boettcher12}%
  \BibitemOpen
  \bibfield  {author} {\bibinfo {author} {\bibfnamefont {I.}~\bibnamefont
  {Boettcher}}, \bibinfo {author} {\bibfnamefont {J.~M.}\ \bibnamefont
  {Pawlowski}}, \ and\ \bibinfo {author} {\bibfnamefont {S.}~\bibnamefont
  {Diehl}},\ }\href {\doibase
  http://dx.doi.org/10.1016/j.nuclphysbps.2012.06.004} {\bibfield  {journal}
  {\bibinfo  {journal} {Nucl. Phys. B}\ }\textbf {\bibinfo {volume} {228}},\
  \bibinfo {pages} {63 } (\bibinfo {year} {2012})}\BibitemShut {NoStop}%
\bibitem [{\citenamefont {Gubbels}\ and\ \citenamefont
  {Stoof}(2011)}]{Gubbels11}%
  \BibitemOpen
  \bibfield  {author} {\bibinfo {author} {\bibfnamefont {K.~B.}\ \bibnamefont
  {Gubbels}}\ and\ \bibinfo {author} {\bibfnamefont {H.~T.~C.}\ \bibnamefont
  {Stoof}},\ }\href {\doibase 10.1103/PhysRevA.84.013610} {\bibfield  {journal}
  {\bibinfo  {journal} {Phys. Rev. A}\ }\textbf {\bibinfo {volume} {84}},\
  \bibinfo {pages} {013610} (\bibinfo {year} {2011})}\BibitemShut {NoStop}%
\bibitem [{\citenamefont {Berges}\ \emph {et~al.}(2002)\citenamefont {Berges},
  \citenamefont {Tetradis},\ and\ \citenamefont {Wetterich}}]{Berges02}%
  \BibitemOpen
  \bibfield  {author} {\bibinfo {author} {\bibfnamefont {J.}~\bibnamefont
  {Berges}}, \bibinfo {author} {\bibfnamefont {N.}~\bibnamefont {Tetradis}}, \
  and\ \bibinfo {author} {\bibfnamefont {C.}~\bibnamefont {Wetterich}},\ }\href
  {\doibase doi:10.1016/S0370-1573(01)00098-9} {\bibfield  {journal} {\bibinfo
  {journal} {Phys. Rep.}\ }\textbf {\bibinfo {volume} {363}},\ \bibinfo {pages}
  {223} (\bibinfo {year} {2002})}\BibitemShut {NoStop}%
\bibitem [{\citenamefont {Delamotte}(2012)}]{Delamotte12}%
  \BibitemOpen
  \bibfield  {author} {\bibinfo {author} {\bibfnamefont {B.}~\bibnamefont
  {Delamotte}},\ }in\ \href {\doibase 10.1007/978-3-642-27320-9_2} {\emph
  {\bibinfo {booktitle} {Renormalization Group and Effective Field Theory
  Approaches to Many-Body Systems}}},\ \bibinfo {series} {Lecture Notes in
  Physics}, Vol.\ \bibinfo {volume} {852},\ \bibinfo {editor} {edited by\
  \bibinfo {editor} {\bibfnamefont {A.}~\bibnamefont {Schwenk}}\ and\ \bibinfo
  {editor} {\bibfnamefont {J.}~\bibnamefont {Polonyi}}}\ (\bibinfo  {publisher}
  {Springer, Berlin/Heidelberg},\ \bibinfo {year} {2012}), pp.\ \bibinfo {pages}
  {49--132}\BibitemShut {NoStop}%
\bibitem [{\citenamefont {Kopietz}\ \emph {et~al.}(2010)\citenamefont
  {Kopietz}, \citenamefont {Bartosch},\ and\ \citenamefont
  {Sch\"utz}}]{Kopietz_book}%
  \BibitemOpen
  \bibfield  {author} {\bibinfo {author} {\bibfnamefont {P.}~\bibnamefont
  {Kopietz}}, \bibinfo {author} {\bibfnamefont {L.}~\bibnamefont {Bartosch}}, \
  and\ \bibinfo {author} {\bibfnamefont {F.}~\bibnamefont {Sch\"utz}},\ }\href
  {\doibase 10.1007/978-3-642-05094-7} {\emph {\bibinfo {title} {Introduction
  to the Functional Renormalization Group}}}\ (\bibinfo  {publisher}
  {Springer},\ \bibinfo {address} {Berlin},\ \bibinfo {year}
  {2010})\BibitemShut {NoStop}%
\bibitem [{\citenamefont {Wetterich}(1993)}]{Wetterich93}%
  \BibitemOpen
  \bibfield  {author} {\bibinfo {author} {\bibfnamefont {C.}~\bibnamefont
  {Wetterich}},\ }\href {\doibase doi:10.1016/0370-2693(93)90726-X} {\bibfield
  {journal} {\bibinfo  {journal} {Phys. Lett. B}\ }\textbf {\bibinfo {volume}
  {301}},\ \bibinfo {pages} {90} (\bibinfo {year} {1993})}\BibitemShut
  {NoStop}%
\bibitem [{not({\natexlab{b}})}]{note5}%
  \BibitemOpen
  \href@noop {} {} \bibinfo {note} {Note that for physical
  quantities defined at scale $k=\Lambda$, the density and therefore the Fermi
  energy $E_F=k_F^2/2m$ are obtained from mean-field theory. In practice, we
  find a small change in the density (corresponding to a change in $k_F$ of
  order 1 percent) when fluctuations beyond mean field are taken into account
  in the RG approach.}\BibitemShut {Stop}%
\bibitem [{\citenamefont {Wetterich}(2008)}]{Wetterich08}%
  \BibitemOpen
  \bibfield  {author} {\bibinfo {author} {\bibfnamefont {C.}~\bibnamefont
  {Wetterich}},\ }\href {\doibase 10.1103/PhysRevB.77.064504} {\bibfield
  {journal} {\bibinfo  {journal} {Phys. Rev. B}\ }\textbf {\bibinfo {volume}
  {77}},\ \bibinfo {pages} {064504} (\bibinfo {year} {2008})}\BibitemShut
  {NoStop}%
\bibitem [{\citenamefont {Floerchinger}\ and\ \citenamefont
  {Wetterich}(2008)}]{Floerchinger08}%
  \BibitemOpen
  \bibfield  {author} {\bibinfo {author} {\bibfnamefont {S.}~\bibnamefont
  {Floerchinger}}\ and\ \bibinfo {author} {\bibfnamefont {C.}~\bibnamefont
  {Wetterich}},\ }\href {\doibase 10.1103/PhysRevA.77.053603} {\bibfield
  {journal} {\bibinfo  {journal} {Phys. Rev. A}\ }\textbf {\bibinfo {volume}
  {77}},\ \bibinfo {eid} {053603} (\bibinfo {year} {2008})}\BibitemShut
  {NoStop}%
\bibitem [{\citenamefont {Dupuis}(2009)}]{Dupuis09b}%
  \BibitemOpen
  \bibfield  {author} {\bibinfo {author} {\bibfnamefont {N.}~\bibnamefont
  {Dupuis}},\ }\href {\doibase 10.1103/PhysRevA.80.043627} {\bibfield
  {journal} {\bibinfo  {journal} {Phys. Rev. A}\ }\textbf {\bibinfo {volume}
  {80}},\ \bibinfo {pages} {043627} (\bibinfo {year} {2009})}\BibitemShut
  {NoStop}%
\bibitem [{\citenamefont {Sinner}\ \emph {et~al.}(2010)\citenamefont {Sinner},
  \citenamefont {Hasselmann},\ and\ \citenamefont {Kopietz}}]{Sinner10}%
  \BibitemOpen
  \bibfield  {author} {\bibinfo {author} {\bibfnamefont {A.}~\bibnamefont
  {Sinner}}, \bibinfo {author} {\bibfnamefont {N.}~\bibnamefont {Hasselmann}},
  \ and\ \bibinfo {author} {\bibfnamefont {P.}~\bibnamefont {Kopietz}},\ }\href
  {\doibase 10.1103/PhysRevA.82.063632} {\bibfield  {journal} {\bibinfo
  {journal} {Phys. Rev. A}\ }\textbf {\bibinfo {volume} {82}},\ \bibinfo
  {pages} {063632} (\bibinfo {year} {2010})}\BibitemShut {NoStop}%
\bibitem [{not({\natexlab{c}})}]{note1}%
  \BibitemOpen
  \href@noop {} {} \bibinfo {note} {In a finite-temperature
  phase transition, critical fluctuations are thermal (classical). Quantum
  fluctuations are not important for the critical behavior and can be ignored
  when one considers the Ginzburg criterion determining the size of the
  critical region.}\BibitemShut {Stop}%
\bibitem [{not({\natexlab{d}})}]{note2}%
  \BibitemOpen
  \href@noop {} {} \bibinfo {note} {Since the transition
  temperature $T_c$ is much smaller than the mean-field transition temperature,
  when estimating the Ginzburg length $\xi_G$ one can use the $T=0$ values of
  $Z_A$ and $\lambda$.}\BibitemShut {Stop}%
\bibitem [{not({\natexlab{f}})}]{note3}%
  \BibitemOpen
  \href@noop {} {} \bibinfo {note} {In the BCS limit, we use
  $Z_A=k_F^3/24\pi^2m\rho_0$, $\lambda=mk_F/4\pi^2\rho_0$ with $\rho_0\sim
  T_c^2$. In the BEC limit, $Z_A=ma/32\pi$, $\lambda=(ma)^3/16\pi$, and
  $\rho=4k_F^3/3\pi m^2a$.}\BibitemShut {Stop}%
\bibitem [{not({\natexlab{g}})}]{note7}%
  \BibitemOpen
  \href@noop {} {} \bibinfo {note} {The transition
  temperature $T_c=0.218 E_F$ corresponds to the BEC temperature of a bosonic
  (dimer) gas with density $n_B=n/2$ and mass $m_B=2m$.}\BibitemShut {Stop}%
\bibitem [{\citenamefont {Giorgini}\ \emph {et~al.}(1996)\citenamefont
  {Giorgini}, \citenamefont {Pitaevskii},\ and\ \citenamefont
  {Stringari}}]{Giorgini96}%
  \BibitemOpen
  \bibfield  {author} {\bibinfo {author} {\bibfnamefont {S.}~\bibnamefont
  {Giorgini}}, \bibinfo {author} {\bibfnamefont {L.~P.}\ \bibnamefont
  {Pitaevskii}}, \ and\ \bibinfo {author} {\bibfnamefont {S.}~\bibnamefont
  {Stringari}},\ }\href {\doibase 10.1103/PhysRevA.54.R4633} {\bibfield
  {journal} {\bibinfo  {journal} {Phys. Rev. A}\ }\textbf {\bibinfo {volume}
  {54}},\ \bibinfo {pages} {R4633} (\bibinfo {year} {1996})}\BibitemShut
  {NoStop}%
\bibitem [{not({\natexlab{h}})}]{note6}%
  \BibitemOpen
  \href@noop {} {} \bibinfo {note} {The result $t_G^+\simeq
  7 n^{1/3}a$ can also be written as $t_G^+\simeq 4.4 n_B^{1/3}a_B$ where
  $n_B=n/2$ and $a_B=2a$ are the density and (mean-field) scattering length of
  the dimers, respectively.}\BibitemShut {Stop}%
\bibitem [{\citenamefont {Haussmann}\ \emph {et~al.}(2007)\citenamefont
  {Haussmann}, \citenamefont {Rantner}, \citenamefont {Cerrito},\ and\
  \citenamefont {Zwerger}}]{Haussmann07}%
  \BibitemOpen
  \bibfield  {author} {\bibinfo {author} {\bibfnamefont {R.}~\bibnamefont
  {Haussmann}}, \bibinfo {author} {\bibfnamefont {W.}~\bibnamefont {Rantner}},
  \bibinfo {author} {\bibfnamefont {S.}~\bibnamefont {Cerrito}}, \ and\
  \bibinfo {author} {\bibfnamefont {W.}~\bibnamefont {Zwerger}},\ }\href
  {\doibase 10.1103/PhysRevA.75.023610} {\bibfield  {journal} {\bibinfo
  {journal} {Phys. Rev. A}\ }\textbf {\bibinfo {volume} {75}},\ \bibinfo
  {pages} {023610} (\bibinfo {year} {2007})}\BibitemShut {NoStop}%
\bibitem [{\citenamefont {{Strinati, G. C.}}\ \emph {et~al.}(2002)\citenamefont
  {{Strinati, G. C.}}, \citenamefont {{Pieri, P.}},\ and\ \citenamefont
  {{Lucheroni, C.}}}]{Strinati02}%
  \BibitemOpen
  \bibfield  {author} {\bibinfo {author} {\bibnamefont {{Strinati, G. C.}}},
  \bibinfo {author} {\bibnamefont {{Pieri, P.}}}, \ and\ \bibinfo {author}
  {\bibnamefont {{Lucheroni, C.}}},\ }\href {\doibase
  10.1140/epjb/e2002-00371-x} {\bibfield  {journal} {\bibinfo  {journal} {Eur.
  Phys. J. B}\ }\textbf {\bibinfo {volume} {30}},\ \bibinfo {pages} {161}
  (\bibinfo {year} {2002})}\BibitemShut {NoStop}%
\bibitem [{\citenamefont {Sun}\ and\ \citenamefont {Leyronas}(2015)}]{Sun15}%
  \BibitemOpen
  \bibfield  {author} {\bibinfo {author} {\bibfnamefont {M.}~\bibnamefont
  {Sun}}\ and\ \bibinfo {author} {\bibfnamefont {X.}~\bibnamefont {Leyronas}},\
  }\href {\doibase 10.1103/PhysRevA.92.053611} {\bibfield  {journal} {\bibinfo
  {journal} {Phys. Rev. A}\ }\textbf {\bibinfo {volume} {92}},\ \bibinfo
  {pages} {053611} (\bibinfo {year} {2015})}\BibitemShut {NoStop}%
\bibitem [{\citenamefont {Gaunt}\ \emph {et~al.}(2013)\citenamefont {Gaunt},
  \citenamefont {Schmidutz}, \citenamefont {Gotlibovych}, \citenamefont
  {Smith},\ and\ \citenamefont {Hadzibabic}}]{Gaunt13}%
  \BibitemOpen
  \bibfield  {author} {\bibinfo {author} {\bibfnamefont {A.~L.}\ \bibnamefont
  {Gaunt}}, \bibinfo {author} {\bibfnamefont {T.~F.}\ \bibnamefont
  {Schmidutz}}, \bibinfo {author} {\bibfnamefont {I.}~\bibnamefont
  {Gotlibovych}}, \bibinfo {author} {\bibfnamefont {R.~P.}\ \bibnamefont
  {Smith}}, \ and\ \bibinfo {author} {\bibfnamefont {Z.}~\bibnamefont
  {Hadzibabic}},\ }\href {\doibase 10.1103/PhysRevLett.110.200406} {\bibfield
  {journal} {\bibinfo  {journal} {Phys. Rev. Lett.}\ }\textbf {\bibinfo
  {volume} {110}},\ \bibinfo {pages} {200406} (\bibinfo {year}
  {2013})}\BibitemShut {NoStop}%
\end{thebibliography}


%

\end{document}